# Efficient Design for the Implementation of Wong-Lam Multicast Authentication Protocol Using Two-Levels of Parallelism

Ghada F. ElKabbany[1] and Heba K. Aslan[2]

[1] Informatics Department, Electronics Research Institute
Cairo, Egypt
Ghada_kabbany@yahoo.com

[2] Informatics Department, Electronics Research Institute
Cairo, Egypt
hebaaslan@yahoo.com

**Abstract**

Group communication can benefit from Internet Protocol (IP) multicast protocol to achieve efficient exchange of messages. However, IP multicast does not provide any mechanisms for authentication. In literature, many solutions to solve this problem were presented. It has been shown that Wong and Lam protocol is the only protocol that can resist both packet loss and pollution attacks. In contrast, it has high computation and communication overheads. In the present paper, an efficient design for the implementation of Wong and Lam multicast authentication protocol is proposed. In order to solve the computation overhead problem, we use two-levels of parallelism. To reduce the communication overhead, we use Universal Message Authentication Codes (UMAC) instead of hash functions. The design is analyzed for both NTRU and elliptic curve cryptography signature algorithms. The analysis shows that the proposed design decreases significantly the execution time of Wong-Lam protocol which makes it suitable for real-time applications.

## 1. Introduction

Group communication can benefit from Internet Protocol (IP) multicast to achieve efficient exchange of messages. IP multicast is a bandwidth-conserving technology that reduces traffic by simultaneously delivering a single stream of information to thousands of recipients [1]. Applications that take advantage of multicast communication include: video conferencing, distance learning, corporate communications, distribution of software, stock quotes and news. Concerning the security of IP multicast, it has two major drawbacks: first, it does not provide any mechanisms for preventing non-group members to have access to the group communication, which is known as the group confidentiality problem. Second, it does not provide any mechanisms to provide authentication of the sender, which is known as multicast authentication problem. In the present paper, we concentrate only on the multicast authentication problem. For the group confidentiality problem, the reader could refer to [2-13].

The multicast authentication is a serious problem. Authenticity means that the recipient could verify the identity of the sender and ensures that the received message comes from the supposed originator. The solutions of group confidentiality problem are based on the fact that all group members share one symmetric key. In case of any member change, the group key must be modified by a group controller and sent securely to the whole group members. A crucial need is to provide authentication for messages received after a key change. For multicast communication, authentication is a challenging problem, since it requires the verification of data originator by a large number of recipients. Assume a group containing $n$ members. A naïve solution is to use a shared symmetric key between the sender and each recipient to calculate different Message Authentication Codes (MACs). Then, the sender appends the calculated MACs to the group message. Upon receiving the message, each recipient ensures the authenticity of the message using the MAC calculated by the key shared between it and the sender. This solution has a high communication overhead since in order to ensure the authenticity of a message $n$ MACs must be appended to it. Another solution is to use the private key of the sender to sign a hash of the entire message. This solution suffers from the high computation and communication overheads since the signature algorithms require large computation and produce large output signatures. The abovementioned solutions do not resist packet loss, since the loss of any packet of the message will cause the inability to authenticate the received packets. This is due to the fact that the MAC or the signatures are calculated over the whole message. Many multicast applications are running over IP networks, in which several packet losses could occur. To solve this problem, the receiver can request retransmission of the lost packets. In multicast communication, different receivers lose different sets of packets, thus retransmission can overload the resources of both the sender and the network. Therefore, multicast authentication protocols must resist packet loss. In order to resist packet loss, one solution is to calculate MAC or signature for every packet. This solution will suffer from a huge amount of communication and computation overheads. In literature, two solutions for providing multicast authentication were proposed: the first is to design more efficient signature schemes. The latter is to amortize signature over several packets.

In literature, it has been shown that Wong and Lam protocol has several advantages over the other multicast authentication protocols. It is the only protocol that can resist both packet loss and pollution attacks under any circumstances. Also, it has no delay at the receiver, since it could authenticate each packet upon receiving it. Therefore, it is suitable for real-time applications. On the other hand, it has high computation and communication overheads. In the present paper, an efficient design for the implementation of Wong and Lam multicast authentication protocol is proposed. In order to solve the computation overhead problem, we use two-levels of parallelism. To reduce the communication overhead, we use Universal Message Authentication Codes (UMAC) instead of hash functions. UMAC algorithm can achieve the same security level as hash functions with lower output length. Other solution to reduce the communication overhead is to use a signature algorithm with a lower output length (e.g. elliptic curve cryptography which has a lower output length compared to NTRU and RSA for the same security level). Therefore, lower communication overhead could be achieved. The design is analyzed for both NTRU and elliptic curve cryptography signature algorithms, and for different values of message size. The analysis shows that the use of parallel systems decreases significantly the execution time of Wong-Lam protocol which makes the proposed design suitable for real-time applications.

The paper is organized as follows: in Section 2, a survey of multicast authentication is detailed. In Section 3, an overview of multiprocessor schemes is given. Then, the proposed design is detailed in Section 4. In Section 5, discussion of results is given. Finally, the paper concludes in Section 6.

## 2. Survey of Multicast Authentication Protocols

For multicast communication, authentication is a challenging problem, since it requires that a large number of recipients must verify the data originator. According to Wong and Lam [14] and Pannetrat and Molva [15], multicast authentication protocols have several requirements that are summarized below:

- **Delay at sender and receiver:** flows that is real-time in nature need fast processing at sender as well as at receiver.
- **Buffering resources:** the number of packets that have to be stored at both the sender and the receiver in order to carry out the authentication process.
- **Robustness:** the ability of the recipient to authenticate the received packet, even in case of losses in the network (since many of multicast applications are running over IP networks, in which several packet losses could occur).
- **Resistance to packet loss:** the ability of the recipient to start authentication at any arbitrary point in the flow.
- **Resistance to pollution attacks:** the ability of the recipient to distinguish between authenticated packets and modified packets.
- **Latency:** the maximum number of packets that need to be received before a packet can be authenticated.
- **Computational cost:** the computational cost of the protocol.
- **Communication cost:** the number of bytes per packets that need to be appended in order to provide multicast authentication.

To solve the multicast authentication problem, two approaches have been proposed: design more efficient signature schemes and amortize the cost of signature over several packets. For the first approach, efficient digital signature schemes have been proposed in [14-18]. Although these schemes overcome the computational problem, they suffer from the communication overhead problem, which makes them impractical for real-time applications. Another solution is to amortize signature over several packets as proposed in [14, 19 and 20]. Early work was done by Gennaro and Rohatgi [19]. The stream is divided into blocks of *m* packets and a chain of hashes is used to link each packet to the one preceding it. Then only the last packet is signed. Although this approach solves the computation and communication overheads problem, it has a major drawback that is, in case of any packet loss, the authentication chain is broken and subsequent packets cannot be authenticated. Many of multicast applications are running over IP networks where several packet losses could occur. Therefore, multicast authentication protocols must resist packet loss. In [20], Golle and Modadugu solve this problem by appending the hash of a packet into two places: the first is in the next packet and the second is in the packet succeeding by *"a"* places and only the final packet is signed. Their solution is based on the property that loss over the Internet occurs in bursts as stated in

[21] and can resist several bursts of a certain number of packets. Other enhancements to the basic scheme were proposed in order to resist a larger burst. Although they solve the problem of loss over networks, it is not clearly stated how the packet containing signature is sent. The lost of the signature packet requires its retransmission several times. In multicast communication, different receivers lose different sets of packets, thus retransmission can overload the resources of both the sender and the network. Furthermore, the communication overhead will increase.

Wong and Lam proposed in [14] another solution to solve the problem of packet loss. In their proposal, the stream is divided into blocks of $n$ packets ($Pac_1$, $Pac_2$, $Pac_3$…, $Pac_{n-2}$, $Pac_{n-1}$, $Pac_n$) and a tree of hashes of degree 2 is constructed as shown in Figure 1. The hashes of the $n$ packets correspond to the leaves of the tree and only the root of the tree needs to be signed. Each parent corresponds to the hash of its children. For example $H_{1-2}$ = hash of ($H_1$ and $H_2$). Fig. 1 shows the tree construction for a block containing eight packets. In order to authenticate any packet, the siblings of each node along its path to the root and the packet signature must be appended. For example, to authenticate $P_5$, the following sequence must be received: $P_5$, $H_6$, $H_{7-8}$, $H_{1-4}$, $H_{1-8}$ and signature on $H_{1-8}$. The receiver calculates $H'_{5-6}$ using $H_5$ and $H_6$. Then, it calculates $H'_{5-8}$ using $H'_{5-6}$ and $H_{7-8}$. Finally, it calculates $H'_{1-8}$ using $H'_{5-8}$ and $H_{1-4}$ and checks that $H'_{1-8}$ equals $H_{1-8}$ using the received signature. If the check is correct, the received packet will be authenticated. Since each packet carries the information required for its authentication; therefore, any packet loss will not affect the ability of the receiver to authenticate packets arrived after the loss. On the other hand, this solution suffers from a high communication overhead, since it requires the appending of $log_2(n)+1$ hashes to each packet.

Perrig et al. proposed in [22-25] efficient solutions for the authentication problem named Timed for Efficient Stream Loss-tolerant Authentication (TESLA) and Efficient Multi-chained Stream Signature (EMSS). TESLA is based on authenticating packets using MACs and revealing the MAC keys after a certain time interval. Although these solutions have low communication and computation overheads, they have a major drawback that they require that the sender and the receivers maintain the synchronization of their clocks. Furthermore, these solution suffer from the several sent of signature packet in case of packet loss.

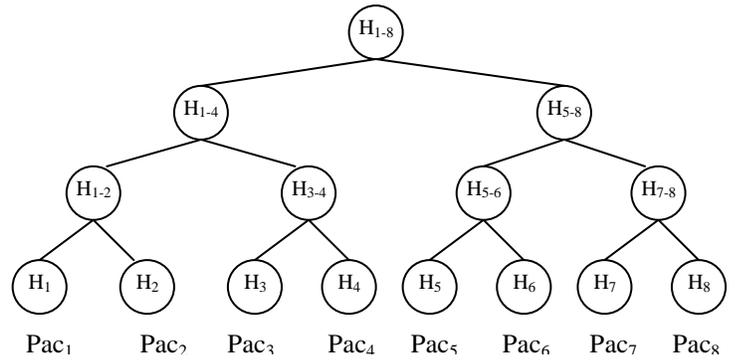

Fig. 1 Tree chaining of the Wong and Lam scheme.

In [15 and 28], a solution was proposed to solve to the problem of multiple sent of signature packet and packet loss using erasure codes. Erasure codes [25-27] allow the receiver to restore the original data under the condition that the loss rate does not exceed a certain value. However, erasure codes can resist only one threat model: packet loss. Erasure codes assume that packets are only lost but not corrupted in transit. Unfortunately, in real environments, packets could be lost, modified, delayed and dropped. These threats are defined in [29] as pollution attacks. In [29], Karlof et al. propose a solution to pollution attacks. In their solution, which is based on the solution given in [28], each symbol output of the erasure code is augmented by additional information – witness information - to differentiate between legitimate symbols and invalid symbols. To obtain witness information, Merkle hash tree is constructed where symbols output of the erasure code are considered as leaves of the tree. Then, each symbol is augmented by the siblings along its path to the root. This information is used to partition symbols as valid or invalid. Then, only valid symbols are used to restore the original packet hashes and the corresponding signature. While this proposal overcomes the pollution attack problem, it has a large communication overhead compared to the abovementioned multicast authentication protocols. In [30 and 31], other solutions to pollution attacks were proposed. They use both public key signature and MAC functions, MAC could have an output that is smaller than that of hash functions [32]. To amortize signature over several packets and resist packet loss, it uses erasure codes. On the other hand, to resist pollution attacks, it uses symmetric key encryption to calculate the witness information instead of the calculation of Merkle hash tree as in [29]. The use of symmetric encryption will lower the communication overhead compared to [29]. In the next section, a background of multiprocessor systems is detailed.

# 3. Why Parallelism?

Many today's advanced research problems need greater computing power at high speeds. Most of these applications are real-time applications which require yielding results at specific deadlines during actual implementation. Parallel systems, which emphasize parallel processing, are the most favorable architectures to increase the computing power and achieve speedup. Parallel processing continues to hold the promise of the solution of more complex problems, by connecting a number of powerful processors together into a single system. These connected processors cooperate to solve a single problem that exceeds the ability of any one of the processors [33]. Depending on how the memory is shared, there exist two models of parallel systems, shared memory systems (tightly coupled) and distributed memory systems (loosely coupled). Shared Memory (SM) systems use a common memory shared by various processors and have a centralized control. This primate shared memory access and the involved processors have overlapping primary address space, which means that one processor can directly access any other processor data. A Distributed Memory (DM) system, involves connecting multiple independent nodes each contains a processor and its local memory. There is no sharing of primary memory, but each processor has its own memory. The contents of each memory can only be accessed by its processor. When a processor needs information owned by another processor, the information is sent as a message from one processor to the other. Messages can carry information between nodes and also synchronization node activities. There are no restrictions on the number of available processors [33]. DM systems have some advantages over SM systems: *First*, in DM systems, as the number of processors increases, the memory size increases, while in SM systems, the memory size does not increase. *Second*, as the number of processors in DM systems increases, the total memory bandwidth increases, while in SM systems, the total memory bandwidth remains constant, independent of the number of processors. *Third*, as the number of processors in DM system increases, processing capability of the system increases. In SM systems, addition of more processors causes memory bottleneck, which decreases processing capability of the system. Also, the number of stages in the network increases with the number of processors. Thus, even if sufficient memory bandwidth can be provided, the minimum network latency increases with the number of processors. *Finally*, the synchronization among processors is required for building scalable parallel computer systems. Synchronization prevents processors from reading results from the memory before other processors write them to the memory. In DM systems, the transmission and reception of messages enforces implicit synchronization among the processors. SM systems require extra support for providing a global shared memory mapped onto the set of distributed local memories [33, 34, 35, 36 and 37]. For the above reasons, the parallel systems based on distributed memory are chosen for our research.

Extraordinary improvements over the past few years in microprocessors, memory, buses, networks and software have made possible the collection of groups of inexpensive personal computers and workstations that in concert have processing power rivaling supercomputers. A cluster is a group of independent nodes which forms a loosely coupled multiprocessor system. Multi–cluster systems enable cost–effective mapping of parallel program tasks into computing hardware. It results primarily from distribution of computing power of a system to parallel programs decomposed into groups of cooperating tasks. The distribution of tasks among clusters improves efficiency of a system due to processor load balancing and better control of the use of communication resources. As a result, cluster computing has recently become a domain of intensive research supported by many practical implementations. In commercially available multi–cluster systems, a cluster contains a set of processors that communicate in most cases through shared memory as shown in Figure 2. Mutual communication among clusters is done by an additional communication network. The standard feature is that internal and external cluster communication is implemented using different hardware which leads to the existence of different latencies. The internal latency is generally much lower than that of external communication. However, another common feature is that the internal cluster communication latency is very sensitive to the cluster size. This is especially visible when communication is based on shared memory where the impeding elements can be a memory bus or single access memory modules [38 and 39]. Therefore, multi-cluster systems combine the advantages of both shared memory and message passing systems. In our work we use multi-cluster system to solve the problem of internal communication. In the next section, description of the proposed design is given.

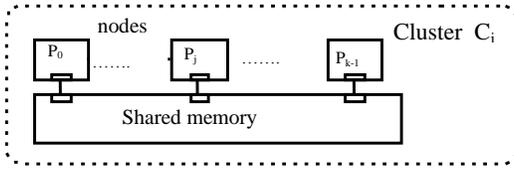

(a) Individual cluster

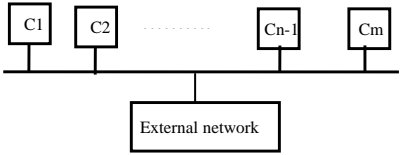

(b) Multi-cluster system

Fig. 2 A Multi-cluster system configuration.

## 4. Efficient Design for the Implementation of Wong-Lam Multicast Authentication Protocol Using Two-Levels of Parallelism

In literature, it has been shown that Wong and Lam protocol has several advantages over the other multicast authentication protocols. It is the only protocol that can resist both packet loss and pollution attacks under any circumstances. Also, it has no delay at the receiver, since it could authenticate each packet upon receiving it. Therefore, it is suitable for real-time applications. On the other hand, it has high computation and communication overheads. In order to solve the computation overhead problem, we use two-levels of parallelism. To reduce the communication overhead, we use Universal Message Authentication Codes (UMAC). UMAC algorithm can achieve the same security level as hash functions with lower output length [32]. Therefore, lower communication overhead could be achieved. Other solution to reduce the communication overhead is to use a signature algorithm with a lower output length (e.g. elliptic curve cryptography which has a lower output length compared to NTRU and RSA for the same security level). In our work, we assume that the data stream is divided into '$G$' groups of packets each group contains '$n$' packets and constructed a tree similar to Figure 1. For the first level of parallelism (coarse grained parallelism), every group of packets is processed in parallel. For '$M$' processors message passing system, there are three cases:

(i) $M > G$: one group of packets is assigned to each processor from '$P_o$' to '$P_{G-1}$' and the other '$M-G$' processors will be idle.
(ii) $M = G$: one group of packets is assigned to each processor.
(iii) $M < G$, we assign $\left\lfloor \frac{G}{M} \right\rfloor$ groups of packets to each processor, and the remaining $\left\{ G - \left\lfloor \frac{G}{M} \right\rfloor * M \right\}$ groups will be assigned to the first $\left\{ G - \left\lfloor \frac{G}{M} \right\rfloor * M \right\}$ processors.

For the first and the third cases, load unbalance may occur. One solution to this problem is to balance the load between processors. This arises the need to divide the group of packets into parts (second level of parallelism - medium grained parallelism). Therefore, more than one processor cooperate to execute the tree hashes. In case of using message passing systems, the communication overhead will increase due to the need to exchange large messages. Using a cluster of processors will solve this problem because more than one processor can share the same memory which results in decreasing the communication overhead. In our work, we assume a multi-cluster system containing '$m$' clusters, where each cluster contains '$k$' processors. Each cluster computes $\left\lfloor \frac{G}{m} \right\rfloor$ group of packets, and the remaining $\left\{ G - \left\lfloor \frac{G}{m} \right\rfloor * m \right\}$ groups will be assigned to the first $\left\{ G - \left\lfloor \frac{G}{m} \right\rfloor * m \right\}$ clusters. Therefore, the maximum number of assigned groups to a cluster '$N_{max}$' is $\left\lfloor \frac{G}{m} \right\rfloor + 1$. $N_{max}$ groups will be executed by '$k$' processors. Each processor computes $\left\lfloor \frac{N_{max}}{k} \right\rfloor$ groups of packets and the remaining $\left\{ N_{max} - \left\lfloor \frac{N_{max}}{k} \right\rfloor * k \right\}$ groups of packets will be executed by the cluster's processors. In order to describe our proposed design, the following parameters are used:

| | |
|---|---|
| $Len_{pac}$ | :packet length (bits) |
| $G$ | :number of groups of packets |
| $InLen_{UMAC}$ | :UMAC input length (bits) |
| $OutLen_{UMAC}$ | :UMAC output length (bits) |
| $Th_{UMAC}$ | :UMAC algorithm throughput (signatures/sec) |
| $Th_{Sig}$ | :signature throughput (signatures/sec) |
| $T_S$ | :total sequential time (msec) |
| $T_{comp}$ | :computation time (msec) |
| $T_{ov}$ | :overhead time (msec) |
| $T_{par}$ | :parallel time/execution time (msec) |

To calculate the computation time for one group of packets '$T_G$', the following equation is used:

$$T_G = T_1 + T_2 + T_3 \qquad (1)$$

Where: $T_1$ is the time to calculate hashes for the first level in the tree and equals to $\left\lceil \dfrac{n * Len_{pac}}{Th_{UMAC}} \right\rceil$, $T_2$ is the time to calculate hashes for the other levels and equals to $\left\lceil \dfrac{(n-1) * InLen_{UMAC}}{Th_{UMAC}} \right\rceil$ and $T_3$ is the signature time and equals to $\left\lceil \dfrac{1}{Th_{Sig}} \right\rceil$. Using Eq. (1), one can calculate $T_s$ which will be equal to $(G * T_G)$. To calculate '$T_{par}$', the following equation is used:

$$T_{par} = T_{comp} + T_{ov} \qquad (2)$$

Where

$$T_{comp} = max\{T_{comp}(P_i)\} \qquad 0 \leq i < M-1 \qquad (3)$$

*and* $T_{ov}$ can be computed using the following equation:

$$T_{ov} = T_c + T_g + T_{app} \qquad (4)$$

Where: $T_c$ is the local communication overhead, which is the time spent on access memory, memory contention and synchronization. $T_g$ is the global communication overhead, which is the time spent on inter-processor communication and $T_{app}$ is the application overhead time, this is the wasted time due to application dependency [36 and 40]. In the next section, analysis of the proposed design for both NTRU and elliptic curve cryptography signature algorithms and different data sizes for Wong-Lam multicast authentication protocol is detailed.

## 5. Evaluation and Experimental Results

The main reason for building parallel computers is to achieve higher performance. The proposed protocol is evaluated for both NTRU and elliptic curve cryptography signature algorithms. In our implementation, we assume that, $Len_{pac}$ is 32 Kbits and the number of packets $n$ is 1024. In addition, for UMAC the $InLen_{UMAC}$ is 128 bits, the $OutLen_{UMAC}$ is 32 bits, $Th_{UMAC}$ is 79.2 Gbps. For NTRU, the output length is 1256 bits and $Th_{sig}$ is 4560 signatures/sec. For elliptic curve cryptography, the output length is 384 bits and $Th_{sig}$ is 5140 signatures/sec.

Figures 3-5 show the analysis of the proposed protocol according to several metrics. Many performance metrics have been proposed to quantify the parallel systems [35-37]. Among of them are:

- Execution time (parallel time) $T_{par}$ is referred to the total running time of the program. The aim of using parallel systems is to decrease the execution time of the problem implementation.
- *Speedup* $S_p$, which relates the time taken to solve the problem on a single processor machine to the time taken to solve the same problem using parallel implementation. $S_p$, of a parallel program running on $M$ processors is defined as the ratio $T_s/T_{par}$. The ideal parallel system (of $M$ processors) will solve the problem $M$ times faster than the serial one ($S_p=M$) and it is said to be linear speedup.
- *Efficiency*, $E_p$, is defined as the ratio $S_p/M$. Optimum computation time, equates to an efficiency of 1 (100%). To achieve this level of efficiency every processor must spent 100% of its time performing useful computation.
- *Degree of improvement* is the percentage of improvement in system performance with respect to sequential execution and can be determined by ($T_s$-$T_{par}$)/$T_s$.
- Finally, *Scalability*, a parallel system is *scalable* if its performance continues to improve as the size of the system (problem sizes as well as the machine size) increase.

Figures 3-5 show the system performance: computation time, speed up, efficiency and the improvement degree. Figure 3 shows the system performance using message passing systems for different message size (1.5 Gbits, 3 Gbits and 5.5 Gbits) where the signature algorithm used is NTRU. Figures 4-5 show the system performance using clustering for a message size equals to 3 Gbits. While in Figure 4, the signature algorithm used is NTRU, in Figure 5, the signature algorithm used is elliptic curve cryptography. From these figures, the following observations are noted:

- From Figures 3-5, it is clear that the use of parallel systems decreases significantly the execution time of Wong-Lam protocol which makes the proposed design suitable for real-time applications.
- From Figures 3(a), 4(a) and 5(a), it has been shown that as the number of processors (clusters) increases, the total execution time decreases irrespective of the message size. This is true for both message passing systems and clustering systems (for both NTRU and elliptic curve cryptography signature algorithms). This leads to the conclusion that the proposed design is scalable.
- From Figures 3(b), 4(b) and 5(b), as the number of processors (clusters) increases, the speedup increases for both message passing systems and clustering systems.

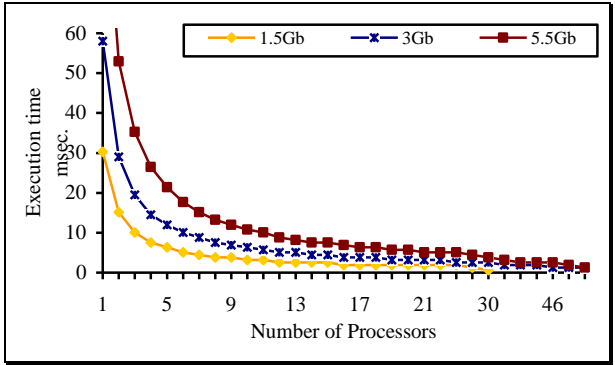
(a) Execution time

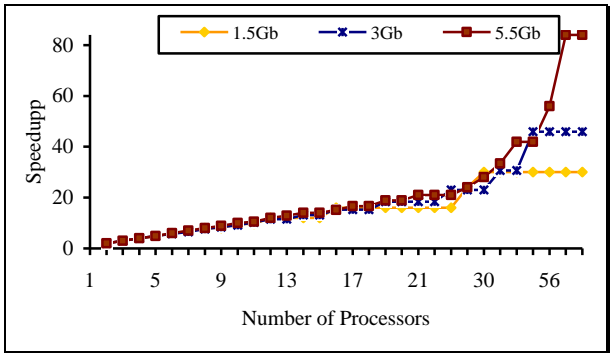
(b) Speedup

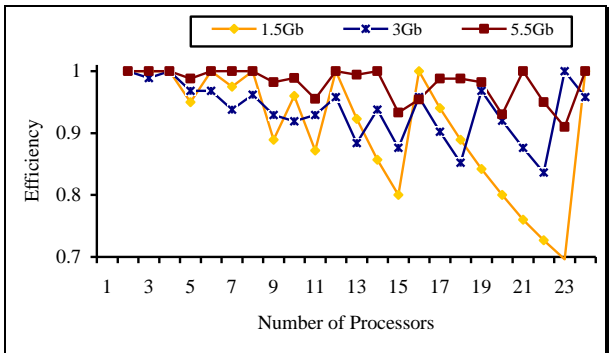
(c) Efficiency

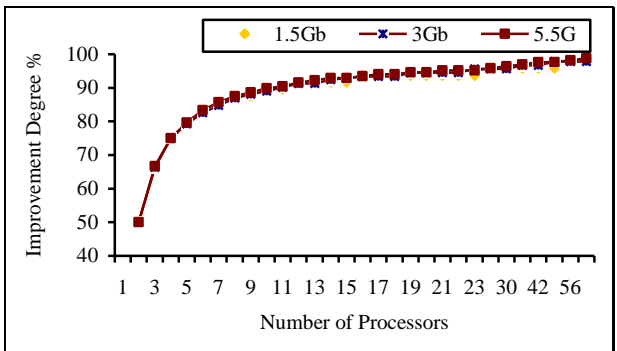
(d) Improvement Degree

Fig. 3 The system performance: execution time, speed up, efficiency and the improvement degree for different message size using message passing systems (the signature algorithm used is NTRU).

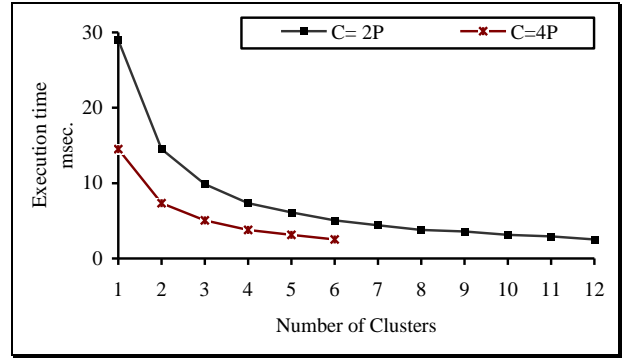
(a) Execution time

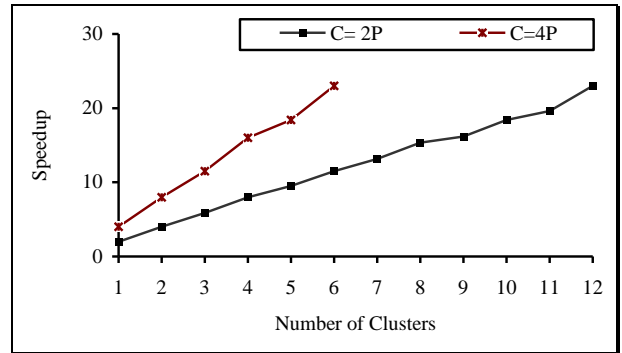
(b) Speedup

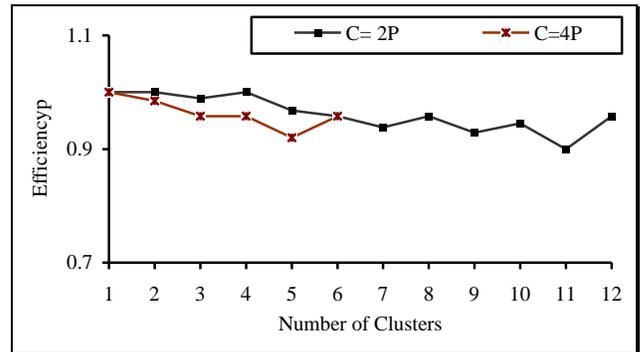
(c) Efficiency

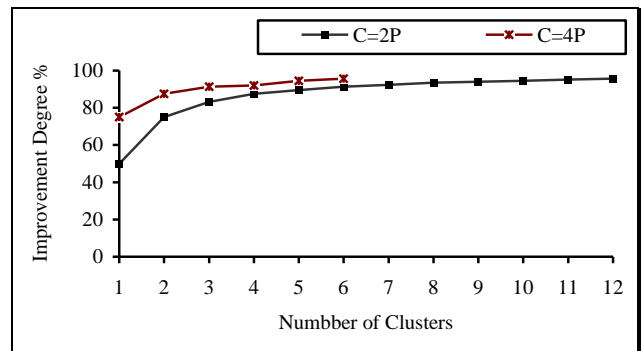
(d) Improvement Degree

Fig. 4 The system performance: execution time, speed up, efficiency and the improvement degree for different cluster size (the signature algorithm used is NTRU and the message size equals to 3 Gbits).

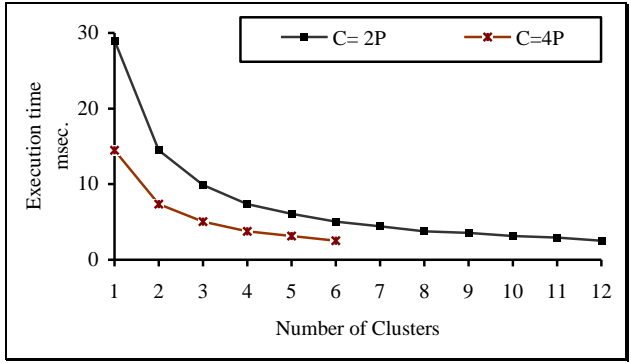
(a) Execution time

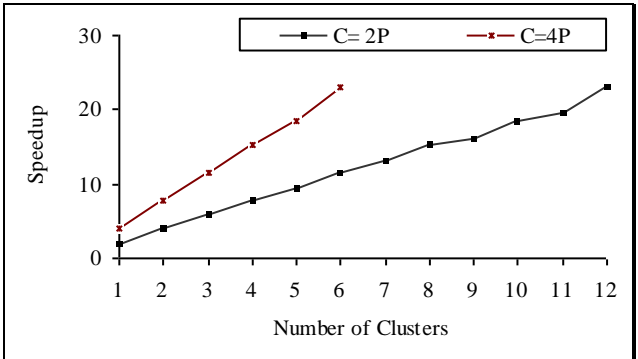
(b) Speedup

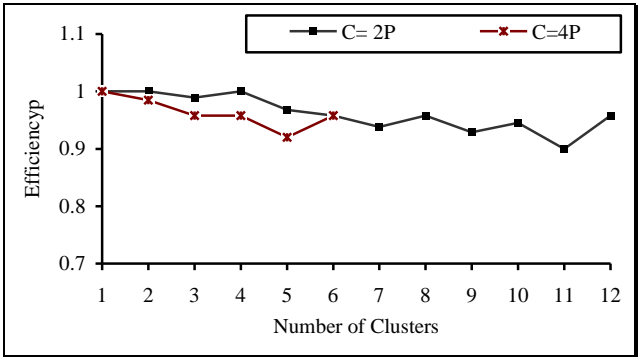
(c) Efficiency

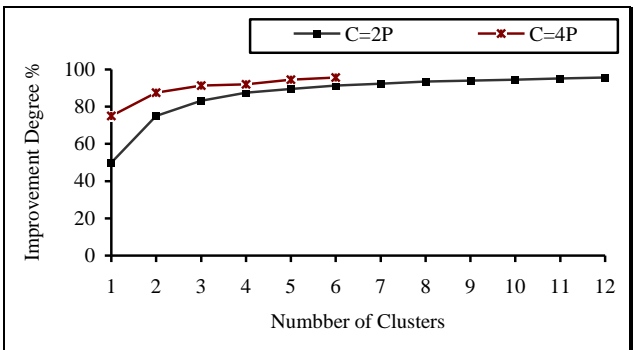
(d) Improvement Degree

Fig. 5 The system performance: execution time, speed up, efficiency and the improvement degree for different cluster size (the signature algorithm used is elliptic curve cryptography and the message size equals to 3 Gbits).

- Figure 3(c) shows that when using message passing systems load unbalance occurs which leads to unstable system efficiency. One solution to this problem is to balance the load between processors. This arises the need to divide the group of packets into parts. Therefore, more than one processor cooperate to execute the tree hashes. In this case the communication overhead will increase due to the need to exchange large messages. Consequently, the use of load balance will increase the total execution time. As a result, we prefer to solve this problem using clustering instead of message passing systems with load balancing as shown in Figures 4 and 5.
- Figures 3(d), 4(d) and 5(d) show the degree of improvement compared to the sequential time. As the number of processors (clusters) increases, the improvement degree increases irrespective of the message size. To obtain a reasonable efficiency, we will be satisfied with an improvement degree equals to 95%. Since, the increase of number of processors will decrease the efficiency.
- From the experimental results, the use of clustering systems enhances the system performance, since it solves the problem of load unbalance. Moreover, for NTRU signature algorithm, the ideal cluster design is a cluster of two processors. This is due to the fact that the time of calculating the tree of hashes is twice the time needed to calculate the signature. On the other hand, for elliptic curve cryptography signature algorithm, a cluster of four processors is the optimum design since the time of calculating the tree of hashes is four times the time needed to calculate the signature.

The analysis shows that the use of multiprocessor system will enhance the system performance. Increasing the number of processors reduces the total execution time. Furthermore, using clustering systems solves the problem of load unbalance.

## 6. Conclusions

In the present paper, the problem of authenticating multicast communication is addressed. Multicast applications are generally running over IP multicast protocol. IP multicast protocol does not have any mechanisms to provide security to transmitted data. The main security features that need to be existed in order to build a secure system are: confidentiality and authentication. In the present paper, we are only concerned about authentication problem. Concerning multicast authentication, the problems confronting providing authentication are detailed. The main attacks that could threaten a multicast authentication protocol are: packet loss and pollution attacks. Two approaches have been proposed to solve the authenticity problem: design more efficient signature schemes and amortize the cost of signature over several packets. In literature, it has been shown that Wong and Lam protocol has several advantages over the other multicast authentication protocols. It is the only protocol that can resist both packet loss and pollution attacks under any circumstances. Also, it has no delay at the receiver, since it could authenticate each packet upon receiving it. Therefore, it is suitable for real-time applications. On the other hand, it has high computation and communication overheads. In the present paper, an efficient design for the implementation of Wong and Lam multicast authentication protocol is proposed. In order to solve the computation overhead problem, we use two-levels of parallelism. To reduce the communication overhead, we use Universal Message Authentication Codes (UMAC) instead of hash functions. UMAC algorithm can achieve the same security level as hash functions with lower output length. Therefore, lower communication overhead could be achieved. Other solution to reduce the communication overhead is to use a signature algorithm with a lower output length (e.g. elliptic curve cryptography which has a lower output length compared to NTRU and RSA for the same security level). The design is analyzed for both NTRU and elliptic curve cryptography signature algorithms and for different values of message size. The analysis shows that the use of parallel systems decreases significantly the execution time of Wong-Lam protocol which makes the proposed design suitable for real-time applications.

**Ghada Farouk Mahmoud ElKabbany** is an Assistant Professor at Electronics Research Institute, Cairo- Egypt. She received her B.Sc. degree, M.Sc. degree and Ph.D. degree in Electronics and Communications Engineering from the Faculty of Engineering, Cairo University, Egypt. Her research interests include High Performance Computing (HPC), Robotics, and Computer Network Security.

**Heba Kamal Aslan** is an Associate Professor at Electronics Research Institute, Cairo-Egypt. She received her B.Sc. degree, M.Sc. degree and Ph.D. degree in Electronics and Communications Engineering from the Faculty of Engineering, Cairo University, Egypt in 1990, 1994 and 1998 respectively. Aslan has supervised several masters and Ph.D. students in the field of computer networks security. Her research interests include: Key Distribution Protocols, Authentication Protocols, Logical Analysis of Protocols and Intrusion Detection Systems.